\documentclass[fleqn,useAMS, usenatbib]{mnras}

\usepackage[dvips]{graphics}
\usepackage{graphicx}
\usepackage{amsmath}
\usepackage{epsfig}
\usepackage[latin1]{inputenc}

\usepackage[T1]{fontenc}
\usepackage{ae,aecompl,url}
\usepackage{multicol}
\newcommand\NS{N\!S}

\newcommand{\appropto}{\mathrel{\vcenter{
  \offinterlineskip\halign{\hfil$##$\cr
    \propto\cr\noalign{\kern2pt}\sim\cr\noalign{\kern-2pt}}}}}

\title[]{Tidal spin down rates of homogeneous triaxial viscoelastic bodies}

\author[Quillen et al.]
{
Alice C. Quillen$^{1,2}$, 
Andrea Kueter-Young$^{1,3}$, 
Julien Frouard$^{4}$, \&  
\newauthor
Darin Ragozzine$^{5}$ 
\\
$^1$Department of Physics and Astronomy, University of Rochester, Rochester, NY 14627, USA \\ 
$^2$\texttt{alice.quillen@rochester.edu} \\
$^3$Physics Department, Siena College, Loudonville, NY 12211, USA, \texttt{ab08kuet@siena.edu} \\
$^4$US Naval Observatory, 3450 Massachusetts Ave NW, Washington DC 20392, USA \\
$^5$Brigham Young University,  Department of Physics and Astronomy, N283 ESC, Provo, UT 84602, USA\\
}

\begin{document}
\maketitle

\begin{abstract}
We use numerical simulations to measure the sensitivity of the tidal spin down rate of a homogeneous triaxial ellipsoid to its axis ratios by comparing the drift rate in orbital semi-major axis to that of a spherical body with the same mass, volume and simulated rheology. We use a mass-spring model approximating a viscoelastic body  spinning around its shortest body axis, with spin aligned with orbital spin axis, and in circular orbit about a point mass.  The torque or drift rate can be estimated from that predicted for a sphere with equivalent volume if multiplied by $0.5 (1 + b^4/a^4)(b/a)^{-4/3} (c/a)^{-\alpha_c}$ where $b/a$ and $c/a$ are the body axis ratios and index $\alpha_c \approx 1.05$ is consistent with the random lattice mass spring model simulations   but $\alpha_c = 4/3$ suggested by scaling estimates.

A homogeneous body with axis ratios 0.5 and and 0.8, like Haumea, has  orbital semi-major axis drift rate about twice as fast as a spherical body with  the same mass, volume and material properties.  A simulation   approximating a mostly rocky body but with 20\% of its mass as ice concentrated at its ends has a drift rate 10 times faster than the equivalent homogeneous rocky sphere. However, this increase in drift rate is not enough to allow Haumea's satellite, Hi'iaka, to have tidally drifted away from Haumea to its current orbital semi-major axis. 
\end{abstract}


\section{Introduction}

The simplicity of analytic formulae for tidal spin down of spherical bodies and the associated drift rate in semi-major axis
 \citep{goldreich63,goldreich68,kaula64,efroimsky09,efroimsky13} 
 has made it possible to estimate tidal spin-down rates  in diverse
 settings, including stars,  exoplanets, satellites, and asteroids (e.g., \citealt{yoder81,efrolainey,ferrazmello08,ogilvie14}).
However, there is uncertainty in how to estimate the spin down rate and associated semi-major axis drift of non-spherical bodies
(though see \citealt{mathis09} for estimates of tidal drift rates for two extended homogeneous bodies)
and this hampers attempts to interpret the dynamical history of systems that include extended spinning bodies like Pluto's  and Haumea's 
satellite systems (e.g., \citealt{ragozzine09,cuk13,weaver16}).

An extended shape might experience enhanced tidal distortion and dissipation
compared to a stronger spherical body.
\citet{ragozzine09} suggested that use of the radius of an object with equivalent volume (the
volumetric radius) in classic tidal formulae
could lead to an underestimate of the tidal torque on 
Haumea.

A  semi-analytical treatment of the gravitational potential inside a triaxial homogeneous body
can give a description of instantaneous internal deformations (or displacements) and stresses
as a function of Cartesian coordinates  \citep{dobrovolskis82}.
If the axes of the tidal force lie along planes of body symmetry, the displacements
depend on 12 dimensionless coefficients that can be computed numerically.
However, analytical computation in the general case is more challenging.
``The reader is cautioned that in the general case where the tide-raising object 
does not lie along a principal axis, as many as 72 unknown coefficients may appear.''
 \citep{dobrovolskis82}.  To semi-analytically compute the tidal torque for a spinning triaxial body
 we would need to numerically compute all these coefficients and then average over body rotation to compute
 secular or average drift rates.

Due to the complexity of accurate semi-analytical computation,
it would be convenient to have scaling relations,  dependent on body axis ratios,
that would allow one to obtain correct 
analytical formulae for tidal evolution from the well known one for homogeneous spherical bodies. 
Our goal here is to numerically measure such correction factors using numerical simulations.

Because of their simplicity and speed, compared to more computationally intensive grid-based or finite element methods, mass-spring computations are an attractive method for simulating deformable bodies.
By including spring damping forces, they can be used to model viscoelastic deformation.
We previously used a mass-spring model to study tidal encounters  \citep{quillen16} and 
measure tidal spin down for spherical bodies over a range
of viscoelastic relaxation timescales  \citep{frouard16}.
Mass spring models
are not restricted to spherical particle distributions and so can be used to study triaxial ellipsoids.
Our work \citep{frouard16} compared orbital semi-major axis drift rates for spherical bodies
to those predicted analytically.  Here we use the same type of simulations
to measure drift rates but work in a setting where analytical computations are lacking and
the numerical measurements may motivate order of magnitude scaling arguments.

\section{Description of mass-spring model simulations}
\label{sec:des}

To simulate tidal viscoelastic response we use the mass-spring model by \citet{quillen16,frouard16}
 that is built on the modular N-body code rebound  \citep{rebound}.  Springs between mass nodes are damped and
so approximate the behavior of a Kelvin-Voigt viscoelastic solid with Poisson ratio of 1/4 \citep{kot15}.
As done previously \citep{frouard16}, we consider a binary in a circular orbit, with a spinning body resolved with masses and springs.  
The other body (the tidal perturber) is modeled as a point mass.
The particles are subjected to three types of forces: the gravitational forces acting on every pair of particles in the body and with a massive companion, and the elastic and damping spring forces acting only between sufficiently close particle pairs. Springs have a spring constant $k$ and a damping rate parameter $\gamma$. 
The number density of springs, spring constants and spring lengths set the shear modulus, $\mu$, whereas
the spring damping rate,
$\gamma$, allows one to adjust the shear viscosity, $\eta$, and viscoelastic relaxation time, $\tau = \eta/\mu$. 
For a Poisson ratio of 1/4, the Young's or elastic modulus $E = 2.5 \mu$. 
The tidally induced spin down rate  is computed  from the drift rate of the semi-major axis measured
in the simulations.  

\citet{frouard16} measured a 30\% difference between the drift rate computed from the simulations
and that computed analytically.   We do not try to resolve this discrepancy here but  instead 
compare the orbital drift rate of homogeneous triaxial ellipsoids to that of a spherical body with the same
mass and volume, assuming
that the cause of the discrepancy is not strongly dependent on body shape or simulated rheology.

We  consider an extended spherical body of mass $M$,
tidally deformed by a point mass perturber of mass $M_*$.  We consider the two body system
with $M$ and $M_*$ in
a circular orbit with orbital semi-major axis $a_o$ and mean motion $n = \sqrt{G(M+M_*)/a_o^3}$. 
The body $M$ spins with spin rate $\sigma=\dot \theta$ and its spin axis is
 orientated parallel to the orbital angular momentum vector.
 Here $G$ is the gravitational constant.  We use $a_o$ to denote orbital semi-major axis
 so as to differentiate it from body ellipsoid semi-major axis, $a$.   We assume that $a_o \gg a$.

For spherical bodies and for the simulations described by \citet{frouard16} we worked with mass in units of $M$, 
distances in units of radius $R$,  time in units of $t_g = \sqrt{R^3/(GM)}$ and
elastic modulus in units of $e_g = GM^2/R^4$ which scales with the 
gravitational energy density or central pressure.

For a non-spherical homogeneous ellipsoid body we could replace the radius 
$R$ with the body's semi-major axis length,  $a$,
or we could use the volumetric radius, $R_v$, the radius of a spherical body with the same
volume.  The volumetric radius was used by  \citet{breiter12} to study the energy dissipation
rate of wobbling and rotating ellipsoids.
 For oblate systems we could use the mean equatorial radius, $R_e=a$.     
We chose the volumetric radius  so that it is straightforward
to compare simulations with the same number of mass-particles and simulated material properties 
(shear modulus, shear viscosity, viscoelastic relaxation time and type of spring network).
For our triaxial ellipsoids we
work with mass in units of $M$, distances in units of volumetric radius $R_v$, 
time in units of
 \begin{equation}
t_g = \sqrt{ \frac{ R_v^3}{GM}} \label{eqn:tgrav} 
\end{equation}
 and elastic modulus $E$ in units of 
\begin{equation}
e_g = \frac{GM^2}{R_v^4}. \label{eqn:eg} 
\end{equation}
Our choice of units implies that $M=1$ and $R_v=1$.
We assume that the ellipsoid spins about a principal body axis,  
about its axis of maximum moment of inertia,
and with spin axis
 orientated in the same direction as the orbital angular momentum vector.

 Our convention for body semi-axis ratios is $a>b>c$ with $c$ oriented along body spin and orbital spin axes.
The volume of a triaxial ellipsoid is $V = abc \frac{4\pi}{3}$
so a sphere with the same volume  has radius of 
\begin{equation} 
R_v = (abc)^\frac{1}{3} = a (b/a)^\frac{1}{3} (c/a)^\frac{1}{3}. 
\end{equation}

\begin{figure}
    \includegraphics[width=3.0in]{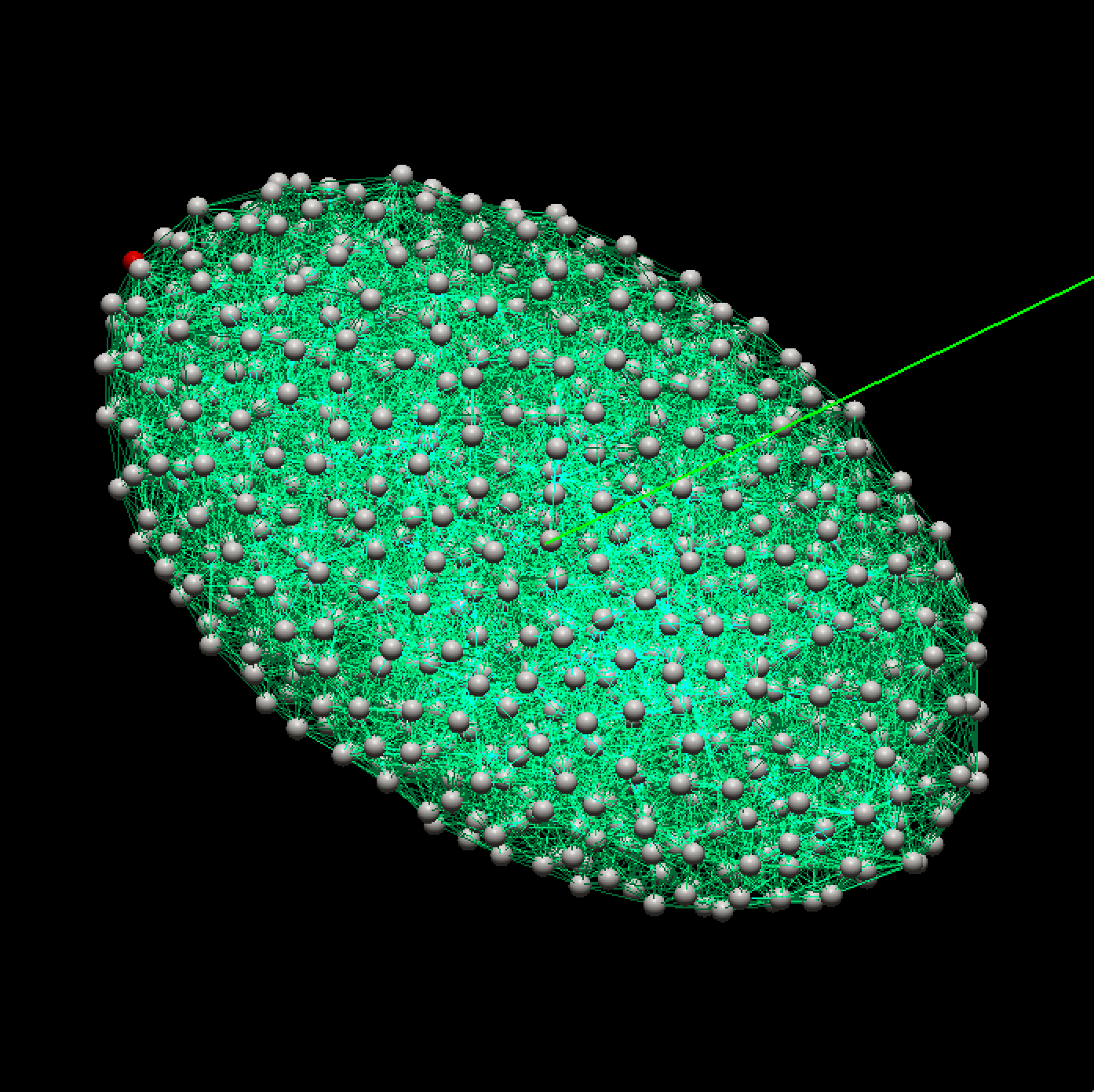}
       \includegraphics[width=3.0in]{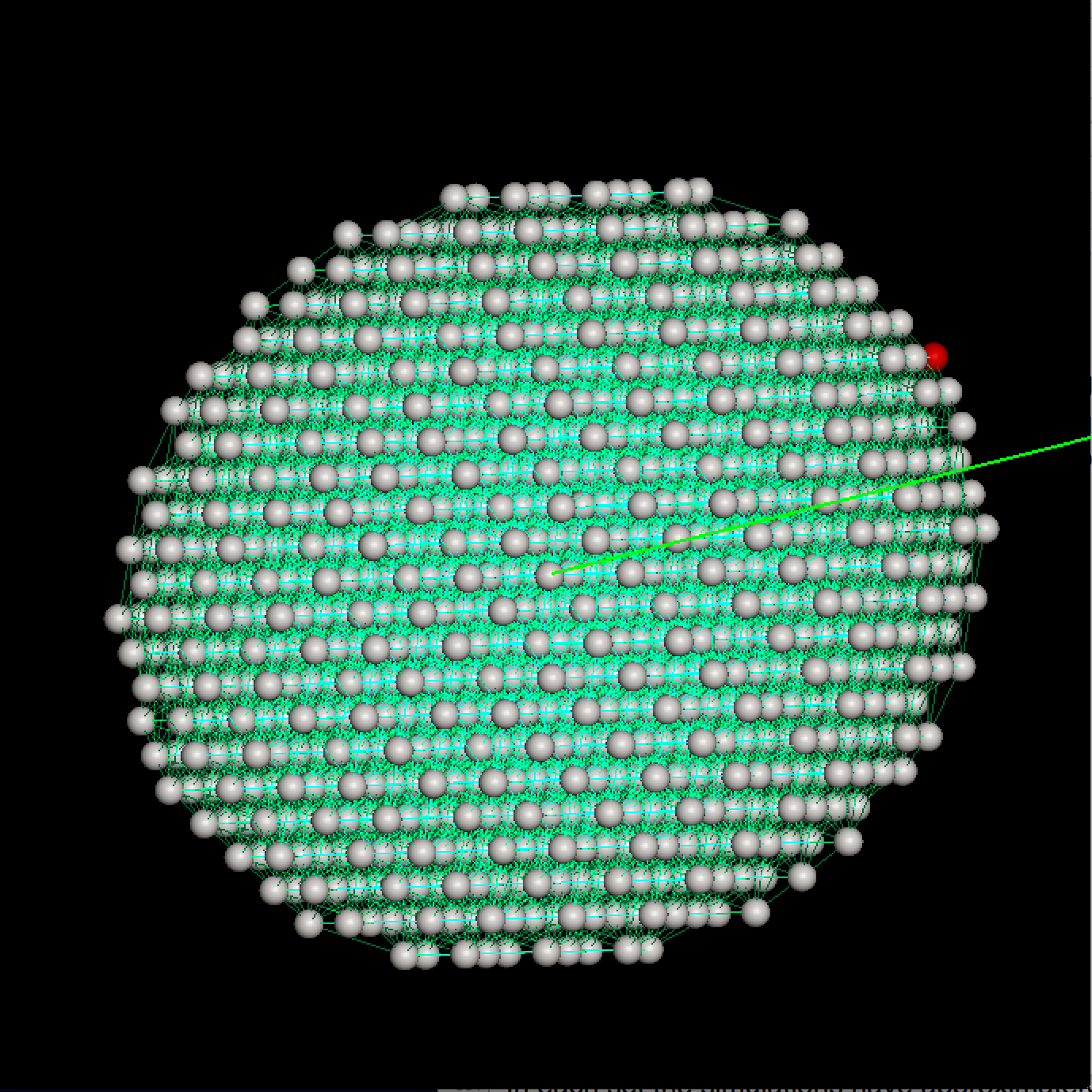}
 \caption{Two simulation snapshots.  The top one shows a triaxial body from the random mass-spring model
 R series
 with $b/a=0.6$ and $c/a=0.5$ as seen looking down on the orbital plane.
 The bottom snapshot shows a cubic lattice simulation for an oblate body with $b/a=1.0$ and $c/a=0.7$.
 This simulation is seen from an inclined angle.
 The long green lines point to the tidally perturbing mass.
 A single mass node is colored red so that body libration can be viewed during tidal lock.
 The mass nodes are rendered as spheres with radius equal to 1/4 the minimum interparticle
 distance $d_I$ and spring connections are shown as thin green lines.  
The mass nodes are kept in their positions
 by spring elastic forces between nodes, not because the spherical surfaces repel.
 \label{fig:snap}}
 \end{figure}

We run two types of mass spring models, a random spring model, 
(as we described previously; \citealt{frouard16}) and a cubic lattice model; for simulation 
snap shots see Figure \ref{fig:snap}.
For the random spring model, particle positions are drawn from an isotropic uniform distribution
but only accepted into the spring network if they are within the surface, 
$x^2/a^2 + y^2/b^2 + z^2/c^2 = 1$,
and if they are more distant than $d_I$ from every other previously generated particle.
The parameter $d_I$ is the minimum inter-particle spacing.
For the cubic lattice model, particle positions are generated in 3-dimensions at Cartesian coordinates, within
the ellipsoidal body surface, that are separated in $x,y$ or $z$ by
$d_I$.  The cubic lattice cell is aligned with the
$x$-axis, the direction between the $M$ and $M_*$,  the $z$-axis, aligned with the orbital
and body spin axes and the $y$-axis  along the tangential orbital motion.
Crystalline lattices are not elastically isotropic as their stiffness depends on the direction
on which stress is applied.
A cubic lattice was chosen because its elastic behavior is the same along $x$, $y$ or $z$ planes.
An advantage of a lattice model is that the spring network is more homogeneous (has less granularity or porosity)
and so can be run with shorter springs and fewer springs per node and this would reduce
the affect of a weaker surface layer (due to fewer springs per node near the surface).
The advantage of the random spring model is that its stiffness should not be sensitive to
the direction on which stress is applied; it should be elastically isotropic.

Once the particle positions have been generated, every pair of particles within $d_s$ of each 
other are connected with a single spring.  The parameter $d_s$ is the maximum rest
length of any spring.  For the cubic lattice we chose 
$d_s$ so that 
cubic cell face diagonals  and cubic cell cross diagonals are connected (see Figure 1 by \citealt{kot15} for an 
illustration).   For the random spring model
we chose $d_s$ so that the number of springs per node is greater than 15, as recommended
by \citet{kot15}.
  
At the beginning of
the simulation the body is not exactly in  equilibrium  because springs are initially set to their rest lengths. 
The body initially vibrates.  As a result
we run the simulation for a time $t_{damp}$ with a higher damping rate $\gamma_{high}$.
We only measure the semi-major axis drift rate using measurements
taken  after the high damping rate is finished and the body is relaxed and in equilibrium.
The body is a permanent triaxial ellipsoid.   
The ellipticity of the body is not due to its rotation.  

The secular part of the semi-diurnal ($l=2$) term in the Fourier expansion of the perturbing
potential (e.g., see appendix by \citealt{frouard16})
gives a torque on a spherical body
\begin{equation}
T = \frac{3}{2a_o} G M_*^2 \frac{R^5}{a_o^5}  k_2 (\omega) \sin \epsilon_2(\omega) 
\end{equation}
(also see \citealt{kaula64,M+D}).
When the inclination and eccentricity are small, conservation of angular momentum gives an estimate of the secular drift rate of the semi-major axis from the secularly averaged torque
\begin{eqnarray}
\frac{\dot a_o}{na_o} & \approx&  - \frac{2 T a_o}{GMM_*} \nonumber \\
&=&
 -3 \left( \frac{M_*}{M} \right) \left( \frac{R}{a_o}\right)^5 k_2 (\omega) \sin \epsilon_2 (\omega)
 \label{eqn:dadt}
\end{eqnarray}
where the tidal frequency $\omega = 2 (n - \dot \theta) = 2 (n - \sigma)$.
The quality function is $k_2  (\omega) \sin \epsilon_2 (\omega)$
and is often approximated as $k_2/Q$ with $Q$ a tidal dissipation factor 
(e.g., \citealt{kaula64}) and $k_2$ a Love number.
At low orbital eccentricity
$\dot a_o$ gains a term that is proportional to the square of eccentricity (cf equation 3
by \citealt{yoder81} based on \citealt{kaula64}).
For low values of $\bar\chi = |\omega \tau| \ll 1$ with $\tau$ the viscoelastic relaxation timescale 
and a stiff homogeneous elastic spherical body
\begin{equation}
\left| k_2  (\omega) \sin \epsilon_2 (\omega) \right| \propto \frac{e_g}{\mu} \bar \chi \label{eqn:quality}
\end{equation}
(taking the low $\bar \chi$ limit of equation 25 by \citealt{frouard16} for the Kelvin-Voigt viscoelastic model)
where $\mu$ is the shear modulus or rigidity.
This is consistent with  $k_2 \propto \rho g  R/\mu \propto GM^2/(\mu R^4)$ in the high rigidity limit for 
a  homogeneous and incompressible elastic
sphere \citep{love27,M+D} and tidal dissipation factor $Q \sim \bar \chi^{-1}$.
Here $\rho$ is density and $g = GM/R^2$ is surface gravitational acceleration.

The semi-major axis drift rate, $\frac{\dot a_o}{na_o}$, is proportional to the tidally induced torque $T$
following from angular momentum conservation.
We compare the  orbital drift rate for bodies with the same mass and volume but different
ellipsoid axis ratios.
Common parameters for the simulations are listed in Table \ref{tab:common}.
Three series are run, the C, R and LR series, with parameters listed in \ref{tab:series}.
C and R series simulations have similar numbers of particles but the C series has a cubic lattice model.
R and LR series are random lattice spring models, with the LR series having more particles than the R series.
Each simulation series has parameters similar or the same as given in Tables \ref{tab:common} 
and \ref{tab:series} but individual simulations within the series have different body axis ratios.

For comparison we measure $\dot a_s$, the semi-major axis drift rate, for a spherical body in the same
series, with values listed in Table \ref{tab:series},
and use it to normalize the semi-major axis drift rates for the non-spherical bodies.
Thus the semi-major axis drift rates for the cubic lattice simulations are divided
by that of the spherical cubic lattice simulation and the semi-major axis drift
rate for the random spring model simulations are divided by that of the similar spherical random simulation.

A fairly low value of the Young's modulus (in units of $e_g$) was used so that the body was soft, 
reducing the integration time required to measure a drift in orbital semi-major axis.
The frequency $\bar \chi$ was chosen to be 0.1   (significantly less than 1) so that we remain in the linear viscoelastic regime (e.g., see \citealt{ferrazmello13,noyelles14,efroimsky15,frouard16})
where the quality function and tidal torque 
are linearly proportional to $\bar \chi$. 
The simulations were run 200 times the period  associated with the semi-diurnal frequency or for a total time
$T_{int} = 200  P_{osc}$ with $P_{osc} = 2 \pi /\omega$.  This length of time is chosen so that we
can average over variations caused by body rotation and compute a secular drift rate in orbital semi-major axis.  
Each simulation in the R and C series required a few hours of computation time on a 2.4 GHz Intel Core 2 Duo from 2010. 

\begin{figure}
    \includegraphics[width=3in]{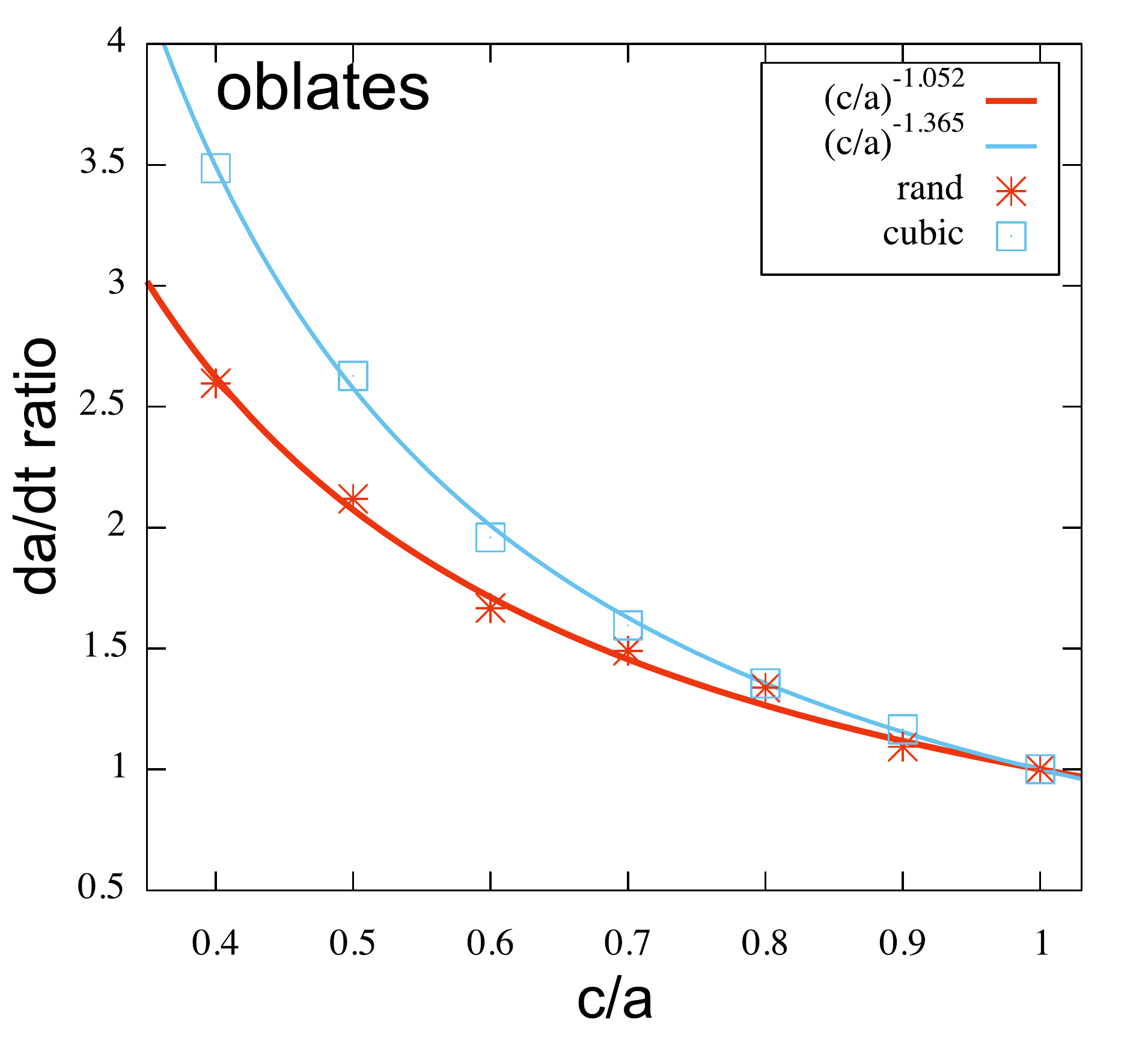}
 \caption{We show  the drift rates in  orbital semi-major axis due to tidal torque as a function of $c/a$ body axis ratio for simulated homogeneous viscoelastic oblate bodies  ($a=b$)  spinning about a minor axis with spin axis
 aligned with the orbital rotation axis.
Measurements from two sets of simulations are shown as points, one based on a cubic lattice, the other
 using the random mass/spring model  (C and R series with parameters listed in Table \ref{tab:series}). 
The simulations in each series have approximately the same numbers of particles, springs per node, body volume and mass, 
 shear modulus, viscoelastic relaxation timescale, initial spin, semi-major axis and perturber mass ratio.
 The differences are in the axis ratio of the simulated oblate ellipsoid. 
The  vertical axis is  $ \dot a/ \dot a_s$ where $\dot a_s$ is that measured for the spherical body in the simulation series.
The horizontal axis shows  body axis ratio  $c/a$.
The curves are power laws with best fitting index listed in the key. 
 \label{fig:oblate}}
 \end{figure}

For both C series cubic lattice and R series random lattice models, we ran simulations of oblate and prolate systems
with axis ratios $c/a = $0.4 to 1 in steps of 0.1.    The normalized orbital semi-major axis drift rates  $\dot a_o/\dot a_s$
are shown in Figures \ref{fig:oblate} and  \ref{fig:prolate} for oblate and prolate bodies.
 For the R series random spring model we also
ran a series of triaxial systems and their drift rates are shown in Figures \ref{fig:tri} and \ref{fig:tri_pr}.
The LR series is similar to the R series but contains more particles and is used to test the accuracy
of the random spring models.  We also ran analogs for Haumea with the LR series. 
In all cases the bodies rotate about their shortest body axis (corresponding to their principal
moment of inertia axis).  


\begin{table}
\vbox to50mm{\vfil
\caption{\large  Common simulation parameters  \label{tab:common}}
\begin{tabular}{@{}lllllll}
\hline
perturber mass     & $M_*$        & 10   \\
orbital semi-major axis   & $a_o$            & 10 \\
mean motion      & $n$               & 0.1 \\
body spin rate     & $\sigma $   & 0.6  \\
integration time for high damping   & $t_{damp}$  &  3 \\
high damping rate & $\gamma_{high}$ & 20 \\
time step           & $dt$   & 0.003  \\
total integration time  & $T_{int}$ & 1260 \\
\hline
\end{tabular}
{\\  The mass and volume of the spinning body are the same for all simulations.
In  our adopted units, $M=1$ and $R_v=1$.  Frequency
$\bar\chi$ is derived from the viscoelastic relaxation time,  
spin frequency $\sigma$ and semidiurnal frequency $\omega$.
Semi-major axis is in
 units of volumetric radius, $R_v$.   The mean motion, body spin rate and damping rates
 are in units of $t_g^{-1}$ (equation \ref{eqn:tgrav}).    Times are in units of $t_g$.
See section \ref{sec:des} for a description of units for the simulations.
%
}
}
\end{table}

\begin{table*}
\vbox to75mm{\vfil
\caption{\large  Parameters for series of simulations  \label{tab:series}}
\begin{tabular}{@{}lllllll}
\hline
Simulation Series            &                  & R  &  C & LR \\
\hline
Particle lattice                 &                   &   random & cubic  & random\\
Young's modulus & $E/e_g$      & 3.1   & 3.1 & 3.0 \\
frequency            & $\bar \chi$    & 0.10 & 0.10 & 0.10 \\
number of mass nodes & $N$              & 1150  & 1240 & 2900 \\
Springs per node &  $ \NS/N$       & 16 & 11 &  15\\
spring constant   & $k$              & 0.06 & 0.1  & 0.0475 \\ 
spring damping  rate   & $\gamma$ & 7.2 & 13 & 15\\  
minimum particle spacing & $d_I$   & 0.135  & 0.15& 0.1 \\ 
maximum spring length & $d_s$ & $2.48 d_I$ & $1.8 d_I$ & $2.38 d_I$\\
\hline
 drift rate for the sphere & $\dot a_s$            & $1.16 \pm 0.06 \times 10^{-6} $ & $1.521 \pm 0.003 \times 10^{-6}$ & $1.45 \pm 0.01 \times 10^{-6}$  \\
\hline
\end{tabular}
{\\  
Numbers of mass nodes, springs, frequency $\bar \chi$ and Young's modulus
are average values for each series.
Parameters $d_s$ and $d_I$ are fixed in each series and used to generate the spring network.
The spring constant $k$ and damping rate $\gamma$ are fixed in each series and set to achieve elastic modulus
$E/e_g \sim 3$ and  frequency $\bar\chi \sim 0.1$.
The drift rate $\dot a_s$ for the sphere in the series is measured from the simulation
output by fitting a line to the orbital semi-major axis
as a function of time.  The error is the rms value of the deviation of the simulation measurements
 from the fitted function.
The damping force depends on the particle mass as by \citet{frouard16} rather than the reduced
mass of the two masses connected by the spring as by \citet{quillen16}.  
Distances ($d_I, d_s$) are in units of volumetric radius $R_v$.  Damping rates are in units of $t_g^{-1}$.
Elastic modulus is in units of $e_g$ (equation \ref{eqn:eg}).
}
}
\end{table*}

 \subsection{Numerical comparisons and tests}
\label{ap:num}

The orbital parameters and mass ratio for our simulations are similar to those used by \citet{frouard16}.  In 
that paper we measured the sensitivity of the predicted to measured orbital semi-major axis drift rate 
 to the size of the initial orbital semi-major axis.
The ratio was independent of orbital semi-major axis, implying that for our adopted semi-major axis of 10
the torque is independent of the higher order terms in the expansion of the tidal gravitational potential.

For the spherical random R series simulations, we ran a comparison simulation
using the adaptive time step 15th-order IAS15 integrator \citep{ias15}. 
The difference between measured drift rates was less than 0.02\% implying that
\texttt{rebound}'s faster and less accurate second order leap-frog integrator for our chosen
step-size is sufficient for our study.

Gravitational interactions are still computed using the computationally intensive but accurate
all particle pairs direct gravity routine in \texttt{rebound} as initial tests with the faster but
less accurate tree-code were not promising. (A soft body that was barely strong enough to
withstand self-gravity using the direct gravity computation imploded with the tree-code).
Even at the semi-major axis of 10 and mass ratio of 10 (see Tables \ref{tab:common} and \ref{tab:series})
the deformations are small and the orbital semi-major axis drift rates, listed for the spheres in Table \ref{tab:series},
are of order $\dot a_s \sim 10^{-6}$.
The gravity computation must be done accurately
 to measure tidal deformation and evolution.  We leave development of tree or multipole gravity
 acceleration methods for future work.
 
Each time we run a simulation with a random spring network,
a new set of particles is generated and this means there are 
 variations in the spring network between simulations.
To estimate the variation in measured semi-major axis drift rates due to differences
in the spring network we ran 3 simulations with identical parameters, each in the R series of
simulations,  for the spherical body and for an oblate body with $c/a=0.5$ and for a prolate body with $c/a=0.5$. 
The standard deviation of $da_o/dt$ computed from each set of three simulations was less than 3\% and smaller
than the differences between the drift rates for simulations with body shapes that differ
in axis ratios by 0.1.  The R series of random lattice simulations has sufficient numbers
of particles that differences in the generated spring networks only cause small variations
in orbital semi-major axis drift rate.

In the random spring model, as particles are never generated outside the ellipsoid
surface, there is a higher probability of generating particles just within the surface
than near planar surfaces embedded within the body.  This means
that the surface is slightly denser than the interior.  As springs are generated
for each pair of particles within $d_s$,  the strength of a region depends on the local particle
density.  
 Because there are more particles per unit volume near the surface,
the surface is stronger than expected taking into account only the reduced
number of springs per node due to the absence of particles outside the ellipsoid. 
A way around this problem is to generate a random distribution of particles in the same manor
but in a larger region that contains our desired ellipsoid and then remove particles
that lie outside the ellipsoid.  We refer to models generated this way as soft-edged
random spring models and those generated with our original procedure as hard-edged
random spring models.  

The hard and soft-edged types of random spring model simulations use the same set of parameters and only
differ in their particle distribution.
The soft-edged distribution is more uniform  but
the surface is more porous and weaker.  Because there are fewer particles near the surface
boundary the density gradient is  shallower near the surface than for the 
hard-edged distribution (that is shown in Figure \ref{fig:snap}).   The softer surface increases
the drift rate, whereas the softer density profile would decrease it.  
We find that the soft-edged bodies have 
higher orbital semi-major axis drift rates so
the softer edge is a more important factor affecting the tidal drift rate.  
For spherical bodies, the difference in drift rate (using the R series of parameters and between
hard- and soft-edged bodies)
is 20\%, however the difference (between hard and soft) in drift rate for the prolate with $c/a=0.5$ is 43\%.
The difference depends on the body surface area and so the prolate models
are more affected by the softer edge.
Because the hard edge somewhat reduces the effect of the soft surface layer and how it
 affects the drift rates, we opted to run the random spring model simulations
using our original method (hard-edged) for generating the random lattice particle distribution.

Even though the maximum spring length is smaller for the cubic lattice simulations (and
so the thickness of the soft region reduced), because the number of springs per node is
lower than for the random spring lattice, the surface layer for the cubic lattice can be quite weak.
We could strengthen the surface layer  by increasing the maximum spring length
(so that there are more springs per node) but this has the effect of increasing
the depth that is weaker than the interior and any advantage of using the cubic lattice model.

We also ran a series of random lattice model simulations with larger numbers of particles
than the R series that we call the LR series.
The LR series has about 2.5 times the number of mass nodes as the R series
and takes about 6 times longer to run as gravity computations are done using
all pairs of mass nodes (direct rather than using a tree code or a multipole algorithm).   
In the LR series we ran
a spherical model, and oblate and prolate models with axis ratio $c/a = 0.5$.
We also ran a triaxial model with $b/a=0.8$ and $c/a=0.5$.
When normalized to the drift rate of the spherical simulation in the same
series we measured a difference between the ratio $\dot a_o / \dot a_s$ computed
with 1200 particles (R series) and those computed with 3000 particles (LR series) that is 
less than 4\% (when divided by the ratio computed with 3000 particles).   There was no trend;
the LR series ratios did not increase or decrease with axis ratio. 
This test suggests that we are running sufficient numbers of particles to ensure 
that the measured drift rates are not strongly sensitive to the structure of
the surface spring network.  However we must keep in mind that the R and LR series
only differ by 2.5 in the number of particles and the maximum spring length (serving
as a skin depth)
is a significant faction of the volumetric radius in both cases ($d_s = 0.33$ for the R series
and 0.24 for the LR series).

\section{Drift Rates Measured for Oblate Bodies}
\label{sec:oblate}

Bodies with $a=b$ and $a>c$ are oblate and here they are spinning about
their short axis.
Figure \ref{fig:oblate} plots as points drift rates measured for the oblate bodies
from our simulations,
normalized to the sphere with the same volume for the random R series
and cubic C series simulations with  parameters listed in Tables \ref{tab:common} and \ref{tab:series}.
With the points we have drawn power law curves  $\dot a_o/\dot a_s = (c/a)^{-\alpha_o}$ with index
$\alpha_o$ that gives the best fit to the numerically measured points.
The index was measured using a 
nonlinear least-squares Marquardt-Levenberg algorithm and its standard deviation
based on the rms value of the deviation of the simulation measurements from the fitted function.
For the cubic lattice the best fit has $\alpha_o = 1.365 \pm 0.008$ whereas
that for the random lattice simulations has $\alpha_o = 1.052 \pm 0.015$.

For an oblate body with spin parallel to the orbital axis and aligned with
the body's axis of symmetry, the shape of the tidally deformed body is independent of
time.   This suggests that we should consider how the drift rate depends on the equatorial radius, $R_e=a$.
For the oblate ellipsoid, the volumetric radius 
\begin{equation}
R_v = a (c/a)^\frac{1}{3} = R_e (c/a)^\frac{1}{3}. \label{eqn:Rv_oblate} 
\end{equation}

We consider the hypothesis that equation \ref{eqn:dadt} is appropriate for our oblate body
but substituting the equatorial radius for the volumetric radius.
Equation \ref{eqn:dadt} gives $\dot a_o /(na_o)$ as a unitless parameter so that it is independent of
time, and we need not correct for the dependence of our unit of time on body radius (equation \ref{eqn:tgrav}).
The shear modulus, viscosity and viscoelastic relaxation timescales are the same for each simulation
as are the body spin rate $\sigma$, semi-diurnal frequency $\omega$ and semi-diurnal frequency
normalized with the viscoelastic relaxation time $\bar \chi$. 
Equation \ref{eqn:dadt} shows that the orbital drift depends on radius to the 5-th power times the quality function.  However equations  \ref{eqn:eg} and \ref{eqn:quality} imply that when $\bar\chi <1$ the quality function
 is inversely proportional to radius to the 4-th power, due to the normalization of the elastic modulus (with $e_g \propto R_v^{-4}$). 
Taking into account the dependence of $e_g$ on radius $R$ we expect
$\dot a_o/(na_o) \propto R$.  This implies that the ratio of the drift rate predicted using a classical
tidal formula and replacing radius with equatorial radius when normalized with that using the volumetric radius is
$\dot a_o/\dot a_s  = (c/a)^{-1/3}$ as $R_v/a = (c/a)^{1/3}$. 
Our numerical measurements are not consistent with this scaling as we measured a power-law index
3 to 4 times larger in magnitude than 1/3.

\section{Scaling estimates for the tidal drift rates of homogeneous triaxial ellipsoids}

To better understand how the tidal torque and associated drift rate in semi-major axis
depends on body axis ratio we estimate the torque on a tidally deformed  triaxial body.
The quadrupole moment of a mass distribution in Cartesian coordinates
\begin{equation}
Q_{ij} = \int \rho (3 x_i x_j - \delta_{ij} r^2) d^3x
\end{equation}
where $\rho$ is the mass density.
For a uniform triaxial ellipsoid with semi-axes $a,b,c$ and in coordinates aligned with body axes
\begin{equation}
{\bf Q} = \frac{M}{5} \left( 
\begin{array}{ccc} 
2a^2 - b^2 - c^2 & 0 & \\
0 & 2 b^2 - a^2 - c^2 & 0 \\
0 & 0 & 2 c^2 - b^2 - a^2 
\end{array}
\right)
\end{equation}
with mass $M = \frac{4 \pi}{3} \rho abc$.
$Q$ is a tensor so we can rotate it; $Q' = R(\theta) Q R(\theta)^{-1}$ with $R(\theta)$ a rotation matrix.
After rotation of the body by angle $\theta$ in the $xy$ plane we transform $Q$ to $Q'$ finding off diagonal term
\begin{equation}
Q'_{xy} =\frac{3}{10} M \sin (2 \theta) (a^2 - b^2) .
\end{equation}

Distant from the object the quadrupolar contribution to the gravitational potential
\begin{equation}
V_2({\bf x}) = G  \frac{Q_{ij}}{2} \frac{ x_i x_j}{r^5}.
\end{equation}
If a mass $M_*$  is located at $x=a_o$,   $y=z=0$ then the $z$ component of the torque on it due
to the quadrupolar force is
\begin{eqnarray}
 T &=&  \left.  M_*  x\frac{\partial V_2}{\partial y} \right|_{x=a_o,y=0,z=0} = \frac{G  M_* Q'_{xy}}{a_o^3}  \nonumber \\
&=& \frac{GM M_*}{a_o^3} 
\frac{3}{10}  \sin (2 \theta) (a^2 - b^2).  \label{eqn:T}
\end{eqnarray}
This is equivalent to an application of MacCullagh's formula for the instantaneous torque exerted by a planet on the permanent figure of an extended satellite.
If $M_*$ causes tidal deformation of $M$ and $\theta$ is a lag angle (due to viscoelastic response)
then this formula can be
used to estimate the torque and associated spin down rate and semi-major axis drift rate
(e.g., see \citealt{M+D}).


A three dimensional version of Hooke's law relating stress applied on three Cartesian coordinates, $\sigma_x, \sigma_y, \sigma_z$ to strain in the three directions
\begin{eqnarray}
\epsilon_x &=& \frac{1}{E} \left[ \sigma_x - \nu (\sigma_y + \sigma_z) \right] \nonumber \\
\epsilon_y &=& \frac{1}{E} \left[ \sigma_y - \nu (\sigma_z + \sigma_x) \right] \nonumber \\
\epsilon_z &=& \frac{1}{E} \left[ \sigma_z - \nu (\sigma_x + \sigma_y) \right]  \label{eqn:hooke}
\end{eqnarray}
where $\nu$ is the Poisson ratio and $E$ the Young's modulus.
More generally a linear relation between stress and strain or Hooke's law in tensor form  is  
$\sigma_{ij} = \lambda  ({\rm tr} \epsilon) \delta_{ij}  + \mu \epsilon_{ij}$  with Lam\'e constants  $\lambda, \mu$.  
In a coordinate system
that diagonalizes the stress and strain tensors, Hooke's law reduces to equations \ref{eqn:hooke}
as off-diagonal terms are zero.

The tidal acceleration on $M$ from distant mass $M_*$ with $M_*$ located at $x=a_o$,   $y=z=0$ is
\begin{equation}
{\bf a}_T \approx \frac{GM_*}{a_o^3}  (2x, -y, -z) 
\end{equation}
taking only the quadrupolar term.

Instead of applying tidal force throughout the body \citep{dobrovolskis82}, 
we roughly approximate it as an instantaneously applied stress
on the body surface.   

We  approximate  the three stresses  as scaling
with the tidal acceleration, ${\bf a}_T$, on the surface, times mass, $M$,  divided by cross sectional area
(from the midplane perpendicular to the direction applied);
\begin{eqnarray}
\sigma_x &\sim&   \frac{GM_*M}{a_o^3} \frac{2a}{bc} \nonumber\\
\sigma_y &\sim& - \frac{GM_*M}{a_o^3} \frac{b}{ac} \nonumber\\
\sigma_z &\sim& - \frac{GM_*M}{a_o^3} \frac{c}{ab}, \label{eqn:stresses}
\end{eqnarray}
where we orient the body with  semi-major axis $a$ along the $x$ axis, $b$ along the $y$ axis and
$c$ along the z axis.
Using Hooke's law (in the form in equations \ref{eqn:hooke}) this gives surface strains of order
\begin{eqnarray}
\epsilon_x &\sim & \frac{GM_* M}{ a_o^3 } \frac{1}{E abc} ( 2a^2 + \nu(b^2+c^2) ) \nonumber\\
\epsilon_y &\sim &\frac{GM_* M}{ a_o^3 }  \frac{1}{E abc} ( -b^2 + \nu(c^2 - 2a^2) ) \nonumber\\
\epsilon_z &\sim &\frac{GM_* M}{ a_o^3 }  \frac{1}{E abc} ( -c^2 + \nu(b^2 - 2a^2) ).  \label{eqn:strains}
\end{eqnarray}

\subsection{Scaling for Oblate bodies}
\label{ap:oblate}

An oblate body with unperturbed semi-major axes has $a=b$ 
and volumetric radius $R_v = a (c/a)^{1/3}$.
Due to the tidal force the body becomes elongated in the equatorial plane with new
semi-major and minor axes $a' \approx a(1 + \epsilon_x)$ and $b' \approx a(1+ \epsilon_y)$.
We note that the moment $Q'_{xy}$ depends on $a'^2 - b'^2 \appropto \epsilon_x - \epsilon_y$.
Inserting $a',b'$ into the formula for the torque  (replacing $a',b'$ for $a,b$ in 
 equation \ref{eqn:T})  and using equations \ref{eqn:strains} for the strains, we find that 
\begin{equation}
 T_o \appropto  \left(\frac{G M M_*}{a_o^3} \right)^2 \frac{1}{E a^2 c} a^4 (1 + \nu).  \label{eqn:To}
 \end{equation}
 We neglect variations in the lag angle $\theta$. 
 Using the torque to estimate the  semi-major axis drift rate (see equation \ref{eqn:dadt}) 
\begin{eqnarray}
 \frac{\dot a_o}{na_o} &\appropto & 
\left( \frac{M_*}{M} \right) \left(\frac{GM^2}{E a^4} \right) \left(  \frac{a}{a_o} \right)^{5} \frac{a}{c}\\
 &=& \left( \frac{M_*}{M} \right) \left(\frac{GM^2}{E R_v^4} \right) \left(  \frac{R_v}{a_o} \right)^{5} \left( \frac{c}{a} \right)^{-\frac{4}{3}}
\end{eqnarray}
where on the first line we are using the equatorial radius $a$
and on the second line the volumetric radius $R_v$ is used for scaling.  
The first line suggests that using the equatorial radius alone in the classic tidal formulae for
spherical bodies would lead to an underestimate of 
the drift rate by the factor $a/c$.
The dependence on $c$ arises because the estimated stresses (equations \ref{eqn:stresses}) and strains $\epsilon_x, \epsilon_y$  (equations \ref{eqn:strains}) depend on the inverse of the cross-sectional area.
More physically, the strains in $x,y$,  are dependent on the stress in the $z$ direction.  

This series of approximations suggests  that the drift rate for oblate bodies
should scale with the
axis ratio $c/a$ to the -4/3 power when normalized to a sphere with equivalent volume 
\begin{equation}
\left(\frac{\dot a_o}{\dot a_s}\right)_{oblate} \approx \left( \frac{c}{a} \right)^{-\frac{4}{3}}.
\label{eqn:scale_ob}
\end{equation}
This approximation is valid when the viscoelastic relaxation time times
the tidal frequency $\bar\chi <1$.
If the tidal frequency is large compared
to the inverse of the viscoelastic relaxation timescale then the scaling would be different as then
 the quality function is not proportional to shear modulus normalized to $e_g$.

\subsection{Comparison of scaling for oblate bodies with the measurements from simulations}

The index -4/3 from equation \ref{eqn:scale_ob} is closer to the index $-1.05$ that we measured for
 the random lattice model (see section \ref{sec:oblate} and Figure \ref{fig:oblate}) 
 than the -1/3 estimated using the equatorial radius alone in 
the classical tidal formula.  
The stronger dependence on $c$ arises because the stresses and strains depend on the body
cross sectional area.

The body surface area is larger for more extreme axis ratio ellipsoids than the equivalent
sphere. The soft surface layer present in the simulations
should increase the drift rates compared to what is expected in the continuum limit. 
We can consider our numerically measured points to be an upper
limit for the value approached with larger numbers of simulated particles as we expect that 
numerically generated surface softness increases the drift rates.
However we must keep in mind that our hard-edged bodies (see discussion in section \ref{ap:num}) 
also have slightly higher density
near the surface than a homogeneous body and we are not sure how this would
have affected the numerical measurements, though we suspect that the weak surface
has a stronger influence on the drift rates.  As our numerical measurements are likely
to be upper limits, we suspect that a more rigorous calculation (better than
 in section \ref{ap:oblate}) would predict a reduced exponent (flatter than 4/3).
 
 The cubic lattice simulations have best fitting index $- 1.365$ and this is closer to 
 the -4/3 power predicted  by equation \ref{eqn:scale_ob}.  However, as we discuss below, 
 our measurements
 for oblate and triaxial bodies imply that the cubic lattice simulations
badly approximate elastically isotropic bodies.

\subsection{Scaling for Triaxial  bodies}
\label{ap:tri}

Orienting the long axis of a triaxial body along the $x$ axis (setting the tide) and using  equation \ref{eqn:strains}
gives us strain values for a tidally deformed triaxial body.
The aligned but tidally deformed body has 
semi-major axis $a' = a(1 + \epsilon_x)$ and $b'  = b(1+ \epsilon_y)$
giving torque (inserting $a'$ and $b'$ for $a,b$ into equation \ref{eqn:T}) 
\begin{equation}
T_a \appropto a^2 - b^2 + 2a^2 \epsilon_x - 2 b^2 \epsilon_y \label{eqn:Ta}
\end{equation}
and we have neglected terms dependent on the Poisson ratio.
The term independent of the strain should average to zero (and we will see that makes
sense below when we estimate the torque for the same body but rotated by $90^\circ$ with
respect to $M_*$).
We compute
\begin{equation}
a^2 \epsilon_x - b^2 \epsilon_y \sim  \frac{GMM_*}{a_o^3 }\frac{1}{Eabc} ( 2 a^4 + b^4).
\end{equation}

We now consider the same body but rotated by $90^\circ$ (as it can be if rotating about the $z$ axis).
If the middle axis of the body  $b$ is oriented along the $x$ axis then 
\begin{eqnarray}
\sigma_x &\sim&   \frac{GM_*M}{a_o^3} \frac{2b}{ac} \nonumber\\
\sigma_y &\sim& - \frac{GM_*M}{a_o^3} \frac{a}{bc} \nonumber\\
\sigma_z &\sim& - \frac{GM_*M}{a_o^3} \frac{c}{ab} 
\end{eqnarray}
(recall $x$ is along the tidal axis)
giving strains
\begin{eqnarray}
\epsilon_x &\sim & \frac{GM_* M}{ a_o^3 } \frac{1}{Eabc} ( 2b^2 + \nu(a^2+c^2) ) \nonumber\\
\epsilon_y &\sim &\frac{GM_* M}{ a_o^3}\frac{1}{Eabc} ( -a^2 + \nu(c^2 - 2b^2) ) \nonumber\\
\epsilon_z &\sim &\frac{GM_* M}{ a_o^3 }\frac{1}{Eabc} ( -c^2 + \nu(a^2 - 2b^2) ).  \label{eqn:strains2}
\end{eqnarray}
For the prolate body oriented with $b$ axis along $x$
we have deformed axes $a' = b(1  +\epsilon_x)$, $b' = a(1+ \epsilon_y)$
and the resulting torque  (replacing $a',b'$ for $a,b$ in equation \ref{eqn:T}) is 
\begin{equation}
T_b \appropto b^2 - a^2 + 2b^2 \epsilon_x - 2a^2 \epsilon_y. \label{eqn:Tb}
\end{equation}
We compute
\begin{equation}
b^2 \epsilon_x - a^2 \epsilon_y \sim  \frac{GMM_*}{a_o^3 } \frac{1}{Eabc}( 2 b^4 + a^4).
\end{equation}
Taking an average of the two torques $T_a$ and $T_b$  we see that the terms
independent of strain cancel.  The torque averaged over rotation (and neglecting angular dependence of
phase angle $\theta$), is approximated from the average of the
two orientations,
\begin{equation}
 T \appropto  \frac{G M M_*}{a_o^3} \frac{GM_* M}{ a_o^3 }\frac{1}{Ea bc} (a^4 +b^4). 
 \end{equation}
The associated drift rate for a triaxial body we estimate as 
\begin{eqnarray}
 \frac{\dot a_o}{na_o} &\appropto & 
\left( \frac{M_*}{M} \right) \left(\frac{GM^2}{E a^4} \right) \left(  \frac{a}{a_o} \right)^{5} \nonumber \\
&& \times
\left( 1 +  \frac{b^4}{a^4}   \right) \left( \frac{b}{a} \right)^{-1} \left( \frac{c}{a} \right)^{-1}\\
 &=& \left( \frac{M_*}{M} \right) \left(\frac{GM^2}{E R_v^4} \right) \left(  \frac{R_v}{a_o} \right)^{5} \nonumber \\
 && \times
\left( 1 +  \frac{b^4}{a^4}  \right) \left( \frac{b}{a} \right)^{-\frac{4}{3}}  \left( \frac{c}{a} \right)^{-\frac{4}{3}} .
\end{eqnarray}

When normalized to the drift rate for an equivalent volume sphere, the triaxial bodies should have drift rates
\begin{equation}
\left(\frac{\dot a_o}{\dot a_s}\right)_{triaxial} \approx
\frac{1}{2}
\left( 1 + \frac{b^4}{a^4}  \right) \left( \frac{b}{a} \right)^{-\frac{4}{3}}  \left( \frac{c}{a} \right)^{-\frac{4}{3}} ,
\label{eqn:scale_tri}
\end{equation}
where the factor of 1/2 is so that when $a=b=c$ the ratio is 1.
The dependence on $1 + \frac{b^4}{a^4}$ arises from averaging over body rotation
whereas the dependence on  $(b/a)^{-\frac{4}{3}} (c/a)^{-\frac{4}{3}}$ is from
 the dependence of tidal stress on cross sectional area.

These order of magnitude scaling estimates do not compute the stress  and strain fields accurately, take
into account rotational deformation  nor
do they appropriately average over body rotation.

\section{Measurements of Drift Rates for Prolate and Triaxial Bodies}
 
\begin{figure}
\includegraphics[width=3in]{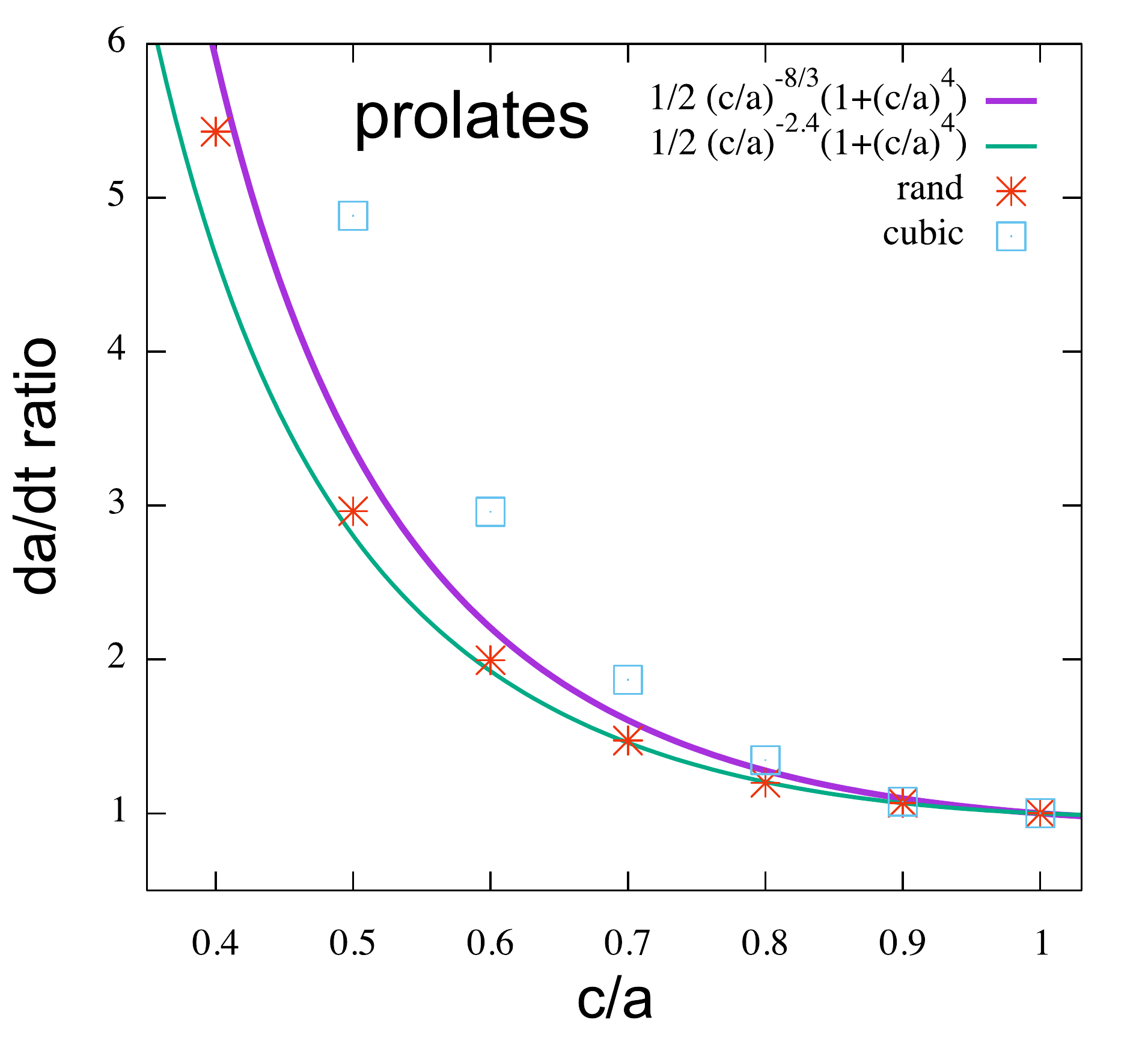}
\caption{We show drift rates in orbital 
semi-major axis as a function of  body axis ratio $c/a$ for simulated viscoelastic prolate bodies 
(with $b=c$) spinning about minor axis and with spin axis aligned with the orbital axis.
Measurements from two sets of simulations are shown as points, one based on a cubic lattice (the C series), the other
 using the random mass/spring model (the R series). 
 The simulations in each series have approximately the same numbers of particles, springs per node, body volume and mass, 
 shear modulus, viscoelastic relaxation timescale, initial spin, semi-major axis and perturber mass ratio.
 The differences are in the axis ratio of the simulated prolate ellipsoid. 
The vertical axis is  $ \dot a/ \dot a_s$ where $\dot a_s$ is that measured for the spherical body in the simulation
series.
The horizontal axis shows body axis ratio $c/a = b/a$.
The drift rates are poorly fit by a power law.
Shown as two curves are equations \ref{eqn:pro1} and \ref{eqn:pro2}. 
Equation \ref{eqn:pro1} is 
based on order of magnitude stress/strain estimates
in section \ref{ap:tri}.  These relations are good matches to the random spring model simulations.
\label{fig:prolate}}
\end{figure}

\subsection{Prolate Bodies}

For the prolate systems, the long axis of the body rotates in the orbital plane and $b=c$.
Figure \ref{fig:prolate} is similar to Figure \ref{fig:oblate} but here we only plot the prolate bodies.  
Power law functions of any index give poor
fits to either cubic or random lattice  prolate simulations. 

For a prolate system with $b=c$ our scaling estimate in equation \ref{eqn:scale_tri} implies that 
\begin{equation}
\left( \frac{ \dot a_o}{a_s} \right)_{prolate} \sim \frac{1}{2}\left( 1 + \frac{c^4}{a^4}  \right) \left( \frac{c}{a} \right)^{-\frac{8}{3}}. \label{eqn:pro1}
 \end{equation}
However if we take into account that we measured a power law dependence of -1.05 for the oblate
systems then we might expect 
\begin{equation}
\left( \frac{ \dot a_o}{a_s} \right)_{prolate} \sim \frac{1}{2}\left( 1 + \frac{c^4}{a^4}  \right) \left( \frac{c}{a} \right)^{-2.4} \label{eqn:pro2}
 \end{equation}
  as $4/3 + 1.05 \approx 2.4$.  We plot both of these curves on Figure \ref{fig:prolate}, finding that both functions
are adequate matches to the prolates from the random spring model simulations but not the cubic lattice ones.
Fitting the curve $\frac{1}{2} \left( 1 + \frac{c^4}{a^4}  \right) \left( \frac{c}{a} \right)^{-\alpha_p}$ we find
a best fitting index of   $\alpha_p = 2.555      \pm    0.017$   for the random spring model 
and $\alpha_p = 3.367        \pm 0.027 $ for the cubic lattice spring model simulations.

At extreme axis ratios, the drift rates for the cubic lattice are much
higher than those of the random lattice (when compared to their matching spherical body).  
The $c/a=0.4$ prolate body cubic lattice model drift rate
has a ratio of $\dot a_o/\dot a_s = 11.6$ and lies above the limits of the plot in Figure \ref{fig:prolate}.
This value is more than twice the value for the random-spring model prolate with $c/a = 0.4$.
The best fitting index for the random spring models $\alpha_p = 2.56$ is near that predicted
from scaling estimates, $8/3 \approx 2.67$. 
However the best fitting index for the cubic lattice model, $3.37$, is much higher
than expected.

In section \ref{ap:tri} we estimated the drift rate by averaging the torque with body orientation
along the tidal axis and that oriented perpendicular to the tidal axis.  
While the strain for the cubic lattice in $x$ and $y$ directions 
(with respect to the orientation of a cubic cell in the lattice)
is the same, the body is weaker when stresses are applied along a direction $45^\circ$ from an edge
of the cubic cell and in a plane containing a cubic cell face. Consequently our
averaging procedure underestimates the average tidal deformation.  
This is likely a stronger affect when the axis ratios are high even when normalizing
to the matching sphere which is also affected by the elastic anisotropy of the cubic lattice.
 The cubic lattice has a shallower
but softer surface (due to a lower number of springs per node) than the random spring model
and this too might contribute to the stronger sensitivity of the drift rate to axis ratio
compared to the random spring model.
We can attribute the high drift rates at low  axis ratios for both oblate and prolate bodies (and a stronger affect
for the prolates) to the anisotropy of the lattice. 

The strong dependence of drift rates on axis ratios emphasizes that the drift rates
are strongly influenced by the weakest part of the simulated body.  When prolate, the body is easiest
to deform along its long axis, when a cubic lattice is present, the drift rate is strongly
influenced by the elastic anisotropy.  Softness in the simulated body surface is likely
our largest source of error in estimating tidally induced drift rates with the random lattice models.
A real asteroid or Kuiper belt object if it contains soft materials, discontinuities or fractures,
may be poorly approximated by a homogeneous strength viscoelastic model.

 \begin{figure}
    \includegraphics[width=3in]{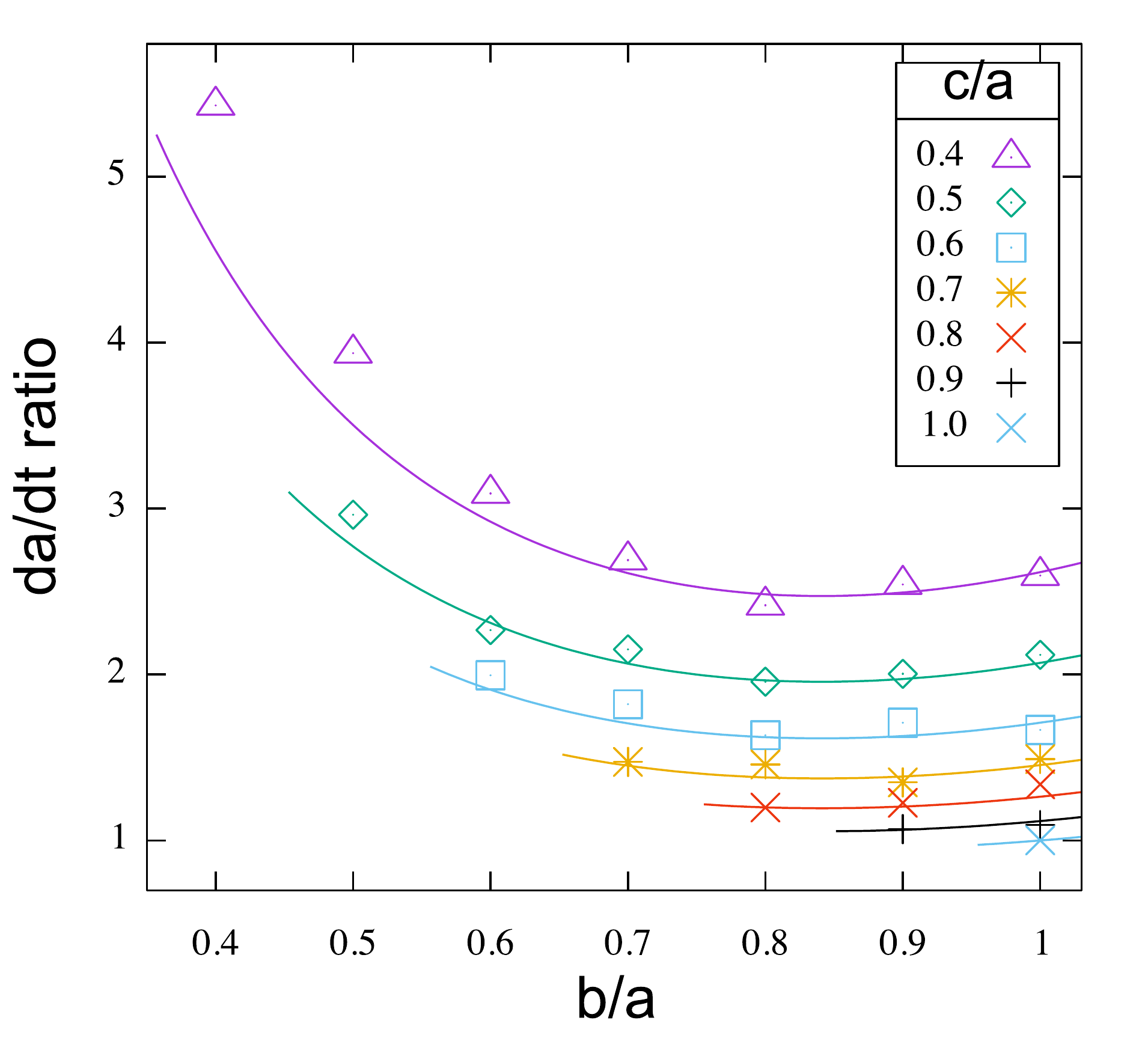}
 \caption{Orbital semi-major axis drift rates for triaxial bodies measured from the random spring model simulations
 from the R series.
 The vertical axis is  $ (\dot a/ \dot a_s)$ where $\dot a_s$ is that measured for the spherical body.
 The horizontal axis shows $b/a$ body axis ratio.
Bodies with different values of $c/a$ have different point types.
 The key on the upper right shows the  body axis ratio $c/a$ for each point type.  
 Oblate bodies lie on the far right.
 The curves show equation \ref{eqn:tri_line}, each line with a different value of $c/a$ and with line color 
  matching the numerically measured points with the same value.
  \label{fig:tri}}
 \end{figure}

\begin{figure}
 \includegraphics[width=3in]{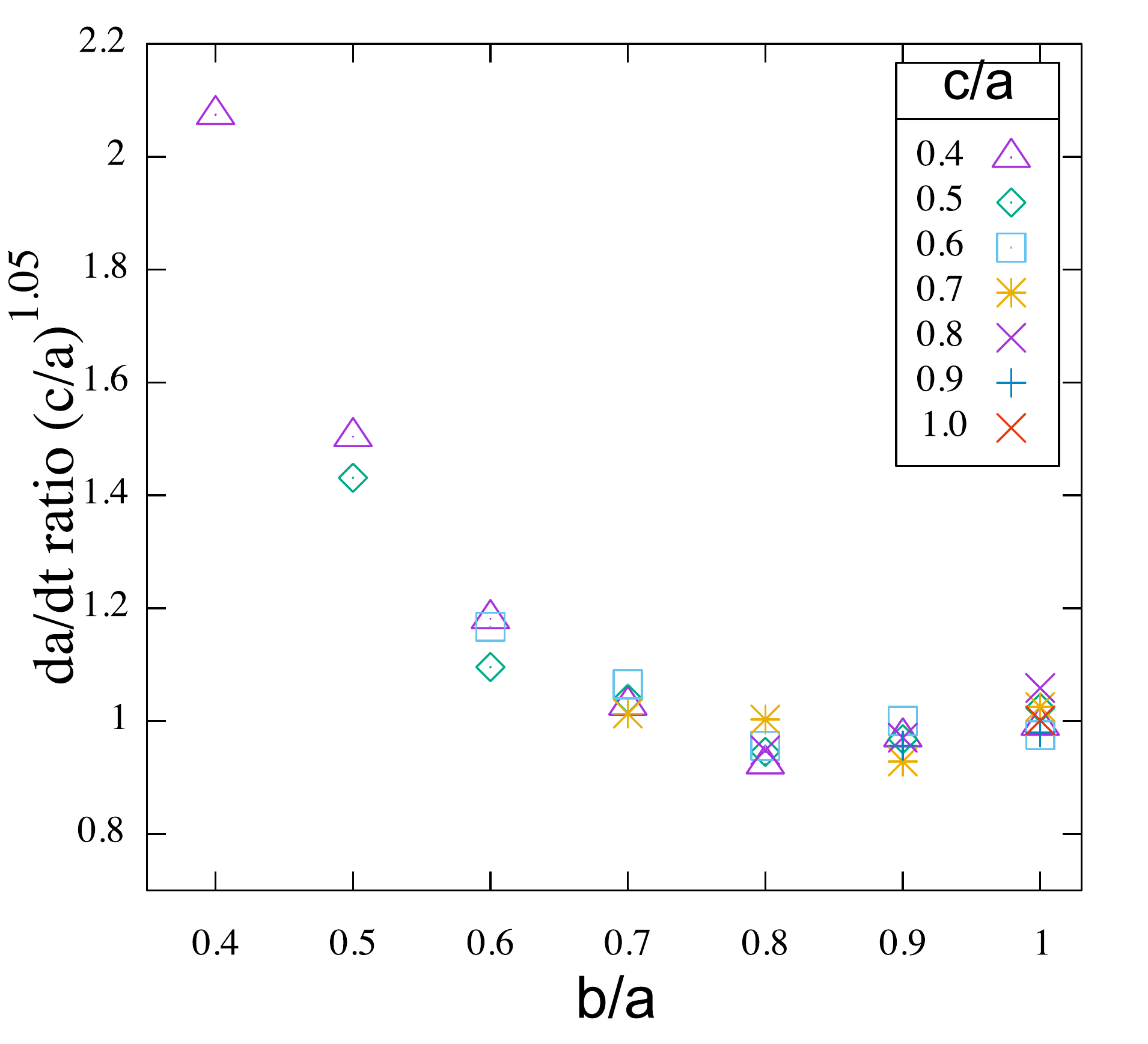}
 \caption{We show the orbital semi-major axis drift rates measured from  the R series of random spring model simulations
 but corrected by $(c/a)^{1.05}$ with power index based on the power law that fit the oblate simulations.
The  vertical axis is  $ \dot a/ \dot a_s \times (c/a)^{1.05}$ where $\dot a_s$ is that measured for the spherical body.
Once corrected for the axis ratio $c/a$, the drift rates for the triaxial bodies resemble those of the prolate bodies.
 \label{fig:tri_pr}}
\end{figure}

\subsection{Triaxial bodies}

Figure \ref{fig:tri} shows (as points) the numerically measured
normalized drift rates for the R series of random spring models
and for triaxial bodies with $c/a = 1.0, 0.9, 0.8, 0.7, 0.5, 0.4$ 
and $b/a$ covering the same range but with $b/a \ge c/a$ so they are stable as
the bodies are rotating about the short axis.  On this figure the point type labels
$c/a$ and the drift rates are plotted versus $b/a$.  The oblate bodies lie on the right
hand side of the plot and the prolate bodies lie to the left of the diagonal connecting upper left
to lower right.
Taking the power law approximations for the oblate bodies we correct the drift rates
by $(c/a)^{1.05}$ and replot the points in Figure \ref{fig:tri_pr}.
In Figure \ref{fig:tri_pr} we see that the points lie on a curve that is similar to
that of the prolate bodies (see Figure \ref{fig:prolate}).
This implies that the drift rate can be considered to be a product of two functions;
one that depends on $c/a$ and is approximately a power law, and the
other that depends on $b/a$.

Our scaling estimate presented in section \ref{ap:tri} suggested
that the drift rate should be a product with form in equation \ref{eqn:scale_tri}. 
That correcting for $c/a$ puts the triaxial drift rates on the same line as the oblates supports the
expectation (based on the order of magnitude estimates)
 that the drift rates can be approximated by a product of functions, one depending
on $c/a$ and the other on $b/a$.

The oblate random lattice models were better fit with a function proportional to $(c/a)^{-1.05}$
rather than $(c/a)^{-4/3}$,
so in Figure \ref{fig:tri} we have plotted on top of the numerically measured points
curves generated with 
\begin{equation}
\left(\frac{\dot a_o}{\dot a_s}\right)_{triaxial} \approx
\frac{1}{2}
\left( 1 + \frac{b^4}{a^4}  \right) \left( \frac{b}{a} \right)^{-\frac{4}{3}}  \left( \frac{c}{a} \right)^{-1.05}. 
\label{eqn:tri_line}
\end{equation}
This function is a pretty good match to all the simulations shown in Figure \ref{fig:tri}.
The slight increase at $b/a$ near 1 is reproduced by both simulations and the estimating
scaling behavior.  Had we plotted curves using equation \ref{eqn:scale_tri}, the curves would have matched
the prolate points on the left but would have diverged from the oblate plots on the right hand side of the plot.
We conclude that the random spring models are probably accurate
and that the function estimated in the section, equation  \ref{eqn:scale_tri},
is a good approximation, though the power-law index for $c/a$ may be somewhat shallower
than -4/3, as given in equation \ref{eqn:tri_line}.

\section{Discussion on Haumea}

\begin{table}
\vbox to90mm{\vfil
\caption{\large  Information about Haumea and Hi'iaka   \label{tab:haumea}}
\begin{tabular}{@{}lllllll}
\hline
Haumea: && \\
Semi-major axis of ellipsoid & $a_H$ & 960 km \\
Axis ratio & $b_H/a_H$ & 0.80 \\
Axis ratio & $c_H/a_H$ & 0.52 \\
Volumetric radius & $R_{vH}$ & 716.6 km \\
Mass of Haumea   & $m_H$ & $4 \times 10^{21}$ kg\\
Energy density scale & $e_{g,H}$ & 4.05 GPa \\
Gravitational timescale & $t_g$ & 1171 s \\
Spin rate  & $\sigma_{H} t_g$ & 0.52 \\
Tidal frequency & $\omega \sim 2 \sigma_{H}$ &  $0.9 \times 10^{-3}$ Hz\\
\hline
Mass ratio & $q = M_{Hi}/M_{H}$ & 0.0045 \\ 
orbital semi-major axis Hi'iaka & $a_{Hi}$ & 49880 km \\
\hline
\end{tabular}
{\\ 
Body semi-major axis and axis ratios are by  \citet{lockwood14}.
Mass of Haumea, mass ratio, $q$, of Hi'iaka and Haumea and semi-major axis are by \citet{ragozzine09}.
The volumetric radius $R_{vH}$ is the radius of a sphere with the same volume as the triaxial ellipsoid.
Spin rate, gravitational timescale, $t_g$ and energy density scale, $e_g$, are computed using the volumetric radius
and equations \ref{eqn:tgrav} and \ref{eqn:eg}.
The spin rate was computed using the spin period  $ P_H = 2\pi/\sigma_H= 3.91531 \pm 0.00005$  hours measured by \citet{lockwood14}.
}
}
\end{table}

The dwarf planet Haumea \citep{brown05} is an extremely fast rotator with
 density  higher than other objects in the Kuiper belt; it is consistent with a body dominated by rock \citep{rabinowitz06,lacerda08,lellouch10,kondratyev16}. 
Visible and infrared light curve fits \citep{lockwood14} find the body consistent with a rapidly rotating oblong
Jacobi ellipsoid shape  in hydrostatic equilibrium 
with axis ratios listed in Table \ref{tab:haumea} (also see \citealt{lellouch10})
and a density of $\rho = 2.6$~g~cm$^{-3}$.
For discussion on formation scenarios for the satellite system see \citet{leinhardt10,schlichting09,cuk13}.
Parameters based on Haumea and Hi'iaka are listed in Table \ref{tab:haumea}.
%

The pressures in the body at depth for a body as massive as Haumea 
would lead to ductile flow giving long-term deformation allowing the body to
approach a figure of equilibrium (a Jacobi ellipsoid).  
 Even if Haumea's shape is  consistent with a hydrostatic equilibrium figure,
on short timescales the body should behave elastically.
For tidal evolution, the relevant tidal frequency is $\omega \sim 2 \sigma_H \sim 10^{-3}$Hz,
comparable to vibrational normal mode frequencies in the Earth.
Our simulations do not allow ductile flow on long timescales, but can approximate the
faster tidal deformations if we model the body as a stiff elastic body with its current shape.

We ran a simulation in the LR random lattice series with axis ratios $b/a=0.8$ and $c/a=0.5$, consistent with
measurements for Haumea.   The LR series of simulations has more particles than the R series 
and is discussed in more
detail  in section \ref{ap:num}.
From the simulation we measure $\dot a_o/a_s = 2.04$  or drift rate
approximately twice that of the equivalent volume sphere.  Equation \ref{eqn:tri_line} predicts a value
2.034, consistent with the numerical measurement, whereas equation \ref{eqn:scale_tri}
gives 2.39.
In their section 4.3.1, \citet{ragozzine09} speculated that using the volumetric radius  
leads to an underestimate of the tidal evolution.  However for the axis ratio of Haumea $b/a \approx 0.8$
and $c/a \approx 0.52$, we find here that the drift rate would only be about twice as fast as estimated using the volumetric radius.  

\begin{figure}
 \includegraphics[width=3.3in]{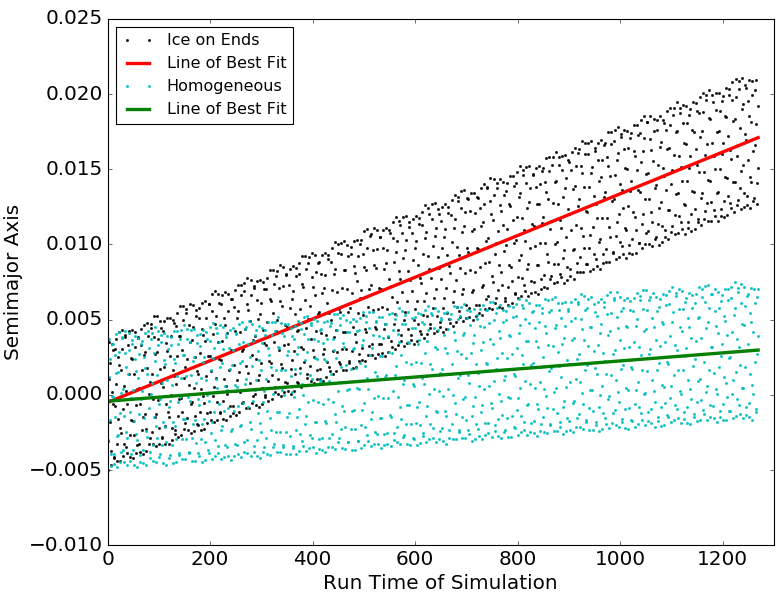}
 \caption{We compare the drift rate in orbital semi-major axis for two simulations
 in the LR series and with axis ratios $b/a=0.8$ and $c/a=0.6$, similar to Haumea.
The blue points show a 
 simulation of a homogeneous body, whereas the black points show a simulation
 with weaker ends (where springs have a  lower spring constant), mimicking a rocky body with icy ends.
 The lines show linear fits measuring the secular drift rate.
 The $x$ axis shows time in units of $t_g$ (equation \ref{eqn:tgrav}) 
 and the $y$ axis shows semi-major axis in units of volumetric radius, $R_v$, measured from the initial value.
The drift rate of the body with soft ends is
about 5.2 times faster than the homogeneous ellipsoid with the same axis ratios and about 10 times faster than
the equivalent volume homogeneous sphere. 
   \label{fig:icy}}
 \end{figure}

\citet{kondratyev16}  proposed that stresses between icy shell and core and associated
relaxation would cause ice to accumulate at the ends of Haumea.  He proposed
that the icy ends could separate forming the two icy satellites Namaka and Hi'iaka.
Estimates for the fraction of ice in a differentiated Haumea range from 7\% \citep{kondratyev16}
to 30\% \citep{probst15}.

 Young's modulus of ice is estimated to be a few GPa (\citealt{nimmo06}; see \citealt{collins10} for a review)
 and this is about 10 times lower than the Young's modulus for rocky materials.
To explore the affect of softer icy ends on the semi-major axis drift rate, we ran
a simulation of a body that is not homogeneous.   Using the same axis ratios of $b/a=0.8$ and $c/a=0.5$
and parameters of the LR series, we reduced
the spring constants to 1/10th the value in the body core at radii greater than 1 (in units of volumetric radius)
 from the body center.
About 20\% of the springs have reduced
spring constants.  This has the effect of lowering the simulated Young's modulus at the
ends of the ellipsoid by a factor of 10.     We did not vary the density as the difference in density between
 ice and rock is much lower than their difference in elastic modulus.
The spring damping parameter $\gamma$ does not vary, so $\tau$, the viscoelastic relaxation 
time-scale, and tidal frequency, $\bar \chi$, are the same in both regions.
The measurements of semi-major axis for this simulation and for the homogeneous one
with the same axis ratios are shown in Figure \ref{fig:icy} along with linear fits that
measure the secular drift rate.
We measured the drift rate in semi-major axis in this simulation, finding that it is
about 5.2 times faster than the homogeneous ellipsoid with the same axis ratios and 10 times faster than
the equivalent homogeneous sphere.   Even a small fraction of softer material
can significantly affect the simulated drift rates.  Perhaps this should have been
expected based on 
the strong sensitivity to elastic anisotropy that we inferred from the cubic lattice model simulations.

The classical tidal formula for the tidal drift rate in semi-major axis
\begin{equation}
\frac{\dot a_o}{n a_o} = \frac{3 k_{2H}}{Q_H} \frac{M_*}{M_H} \left( \frac{R_{vH}}{a_o} \right)^5
\end{equation}
(e.g., \citealt{M+D})
for perturbing object $M_*$ (here Hi'iaka) due to tidal dissipation in the spinning body $M_H$,
where $k_{H}$ and $Q_H$ are the Love number and dissipation factor for Haumea.
We can describe 
corrections to the tidal drift rate by multiplying the right hand side by  a parameter $f_{corr}>1$. 
The above equation implies that $\dot a_o \propto a_o^{-5.5}$ (taking into account dependence on $n a_o \propto a_o^{-1/2}$).
We integrate the above equation for Haumea to estimate the time it takes Hi'iaka to tidally drift outwards to its
 current semi-major axis.
Putting unknowns on the left hand side
\begin{equation}
\frac{k_{2H}}{Q_H} f_{corr} \sim \frac{1}{n \tau_a}  \left( \frac{a_{Hi}}{R_v} \right)^5 \frac{2}{39} q^{-1} \sim 0.1
\end{equation}
with mass axis ratio $q =M_{Hi}/M_{H}$ and where we have used values from
Table \ref{tab:haumea} and an age $\tau_a =  4$ Gyr for the timescale over which tidal migration is taking place.
Here $n$ and $a_{Hi}$ are the mean motion and orbital semi-major axis of Hi'iaka at its current location.
If  $k_{2H}$ for Haumea is as large as 0.01 (at the border of what
would be consistent with rigidity for rocky material) 
and we use corrections for shape and composition
increasing the drift rate by $f_{corr} = 10$ then 
we have equality only if the dissipation parameter is large; $Q \sim 1$.
We conclude that it is unlikely that Hi'iaka alone  tidally drifted to its current location even
if the tidal drift rate is larger by a factor of 10 than estimated using the equivalent volume rocky sphere.

One explanation for the origin of Haumea's satellites and compositional family is a collisional disruption of a past large moon of Haumea  \citep{schlichting09}.   The `ur-satellite' would have formed closer to Haumea
and because of its large mass, could have migrated more quickly than Hi'iaka outward during
the lifetime of the Solar system.  
The failure of our enhanced tidal drift rate estimate to account for Hi'iaka's current position
would suport the `ur-satellite' proposal  (also see discussion by \citealt{cuk13}).

\section{Summary and Discussion}

Motivated by the discovery of spinning elongated bodies such as Haumea, we have carried out
a series of mass-spring model simulations  to measure the tidally induced drift rate (in orbital semi-major
axis) of homogeneous spinning viscoelastic triaxial ellipsoids in a circular orbit about a point mass.
We have restricted this initial study to bodies spinning
about the shortest principal body axis aligned with the orbital axis and with tidal frequency times the
viscoelastic timescale $\bar \chi  \ll 1$, 
 sufficiently small to ensure that the torque is linear in $\bar \chi$.

Our simulations and 
order of magnitude estimates show that the tidal torque or associated orbital semi-major axis drift rates,
when normalized by that of a spherical body of equivalent volume, are described by 
\begin{equation}
\frac{\dot a_o }{\dot a_s} \approx
\frac{1}{2} \left(1 + \frac{b^4}{a^4}\right) \left(\frac{b}{a}\right)^{-\frac{4}{3}} \left( \frac{c}{a} \right)^{-\alpha_c}
\end{equation}
with $\alpha_c \approx 1.05$ consistent with our random lattice simulations but $\alpha_c = 4/3$ predicted
via order of magnitude estimates.
This function is a  good match to the prolate simulations using either value of 
$\alpha_c$ but better matches the oblate and all the triaxial ones with $\alpha_c \approx 1.05$. 

For a homogeneous body with axis ratios equal to those of Haumea ($b/a\approx 0.8$, $c/a \approx 0.5$)
we estimate that the drift rate in orbital semi-major axis  is about twice as fast as that estimated for a spherical body
with the same mass and volume. Motivated by the proposal that ice could have accumulated 
at Haumea's  ends \citep{kondratyev16} we also ran a simulation of a non-homogeneous body with
20\% of the springs  (those at the ends of the body) 
set at 1/10th the strength of those in the core, approximating a body comprised of two materials, ice and rock. 
This simulation has a drift rate 10 times higher than the equivalent homogeneous sphere.
Reexamining the tidal evolution of Hi'iaka, we find that even this increase by 10 is
insufficient to have allowed Hi'iaka to have drifted tidally to its current
location via tidal interaction with Haumea alone.  We have only considered
the behavior of a solid body with fixed axis ratios and a static viscoelastic rheology and we have
 neglected the role of Namaka.
More complex models, perhaps taking into account how material properties and their distribution
are affected by the tidally induced heat, could reexamine this conclusion. 

We experimented with
using a cubic lattice distribution for simulated mass nodes, but suspect that the random spring model is more accurate
because it is elastically isotropic, even though the cubic lattice is more homogeneous and can be
set up with shorter
springs at the same number of mass nodes.  The random spring model is hampered by a soft and weak surface
region with depth set by the maximum length of the springs.   Until we speed up the gravity computation
(perhaps using a multipole method) we cannot on a single processor
increase the number of particles past a few thousand so as
to reduce the effect of the soft surface layer.

Spin orbit resonances and vibrational modes have been neglected from this study and the order of magnitude
scaling estimates rely on crude approximation for the stresses associated with
tidal acceleration.  The mass spring model approximates a Kelvin-Voigt viscoelastic rheology with Poisson
ratio of 1/4
rather than an incompressible Maxwell or Andrade rheology.
Recently developed methods (e.g., \citealt{wisdom08,mathis09,panou14}) might be modified or 
extended to improve upon the scaling arguments presented here.
Future work, both analytical and numerical, will be required to improve upon the 
accuracy of tidal computations for bodies with extreme axis ratios.



\vskip 1.5truein

We are grateful to Valery Lainey, Dan Scheeres,  Dan Tamayo
 and Michael Efroimsky for helpful discussions and correspondence.
This work was improved with helpful and encouraging comments from the referee, Beno\^it Noyelles.
 This work was in part supported by the NASA grant NNX13AI27G and  NSF award PHY-1460352.


{}

\end{document}